\documentclass[journal=jpccck,manuscript=article,layout=twocolumn]{achemso}

\usepackage[T1]{fontenc} % Use modern font encodings

\usepackage{xcolor}
\usepackage{amsmath}
\usepackage{graphicx}

\author{Javier D. Fuhr}
\affiliation{Centro At{\'o}mico Bariloche, CNEA y CONICET, Av. Bustillo 9500, (8400) S.C. de Bariloche (RN), Argentina.}
\email{xx@yy}

\author{Polina M. Sheverdyaeva }
\affiliation{Consiglio Nazionale delle Ricerche, Istituto di Struttura della Materia (CNR-ISM), Strada Statale 14 km 163.5, 34149 Trieste, Italy}

\author{Paolo Moras }
\affiliation{Consiglio Nazionale delle Ricerche, Istituto di Struttura della Materia (CNR-ISM), Strada Statale 14 km 163.5, 34149 Trieste, Italy}

\author{J. Esteban Gayone}
\author{Hugo Ascolani}
\email{hugoascolani@integra.cnea.gob.ar}
\affiliation{Instituto de Nanociencia y Nanotecnolog{\'\i}a (CNEA - CONICET), Nodo Bariloche, Av. Bustillo 9500, (8400) S.C. de Bariloche (RN), Argentina.}

\title{Spin polarization enhancement in a single-layer Bi$_{1-x}$Sb$_x$  alloy on Ag(111) via isovalent substitution}
\begin{document}

\begin{abstract}
Co-adsorption of Bi and Sb on Ag(111) at room temperature yields a single-layer 
Bi$_x$Sb$_{1-x}$ alloy with a rectangular $(3\times\sqrt{3})$ structure containing 
four atoms per unit cell (2/3 ML total coverage) and lacking long-range chemical 
order. We present an electronic structure study of this system combining 
angle-resolved photoemission spectroscopy (ARPES) and density functional theory 
(DFT) calculations. To investigate the effect of inversion symmetry breaking 
induced by substituting a heavier atom (Bi) with a lighter isoelectronic one (Sb) 
within a fixed crystallographic framework, we focused on a Bi-rich composition. 
ARPES measurements reveal four surface-state bands, in good agreement with DFT 
calculations based on a rectangular four-atom overlayer unit cell. DFT calculations 
further show that Sb incorporation induces both in-plane and out-of-plane 
asymmetries in the electronic potential, leading to sizable spin splitting and spin 
polarization of the overlayer bands. Although these effects are partially reduced by 
interaction with the substrate, they remain significant. Our work illustrates, 
through a concrete model system, a general principle: incorporating a lighter 
isovalent element can significantly enhance spin polarization, potentially offering 
a useful design guideline for understanding and engineering Rashba-related systems.
\end{abstract}

\section{\label{intro}Introduction}

Two-dimensional (2D) materials provide a compelling platform for investigating electronic phenomena that emerge from reduced dimensionality and strong spin--orbit coupling (SOC). In 2D systems where SOC is strong and inversion symmetry is broken, spin degeneracy of the electronic bands is lifted, giving rise to Rashba-split surface states \cite{Rashba1983,OAMPRL2011,OAMPRB2013}. This effect has attracted considerable attention over the past decade, as spin-textured electronic structures are of direct relevance to spintronic applications \cite{Bihlmayer2022}.

Sustained research efforts are directed toward both deepening the fundamental understanding of this phenomenon and identifying materials that exhibit giant Rashba splitting. Of particular interest for spin-transfer torque (STT) applications are systems displaying strong out-of-plane spin polarization, that is, spin polarization along the surface normal. To this end, considerable attention has been devoted to relatively simple model systems, including the BiCu$_2$/Cu(111) \cite{BiCu2alloy2020} and SbAu$_2$/Au(111) \cite{JWells2025} surface alloys and one-dimensional Bi chains on Ag(100) \cite{PolinaPRB2022}.

In this context, the bismuth-antimony (Bi$_x$Sb$_{1-x}$) alloy represents an ideal candidate for exploring stoichiometric control of the spin texture at the atomic scale. Both elements are group-V semimetals that naturally tend to form chemically disordered alloys, and since Bi and Sb possess high atomic numbers (83 and 51, respectively), strong spin-orbit coupling (SOC) is expected in their compounds. Indeed, bulk bismuth-antimony exhibits a topological insulator phase within a narrow range of stoichiometries \cite{Hsieh2008,Lemaitre2022}.

In its two-dimensional form, however, the Bi$_x$Sb$_{1-x}$ alloy had not been synthesized until recently, when we demonstrated that high-quality single-layer Bi$_x$Sb$_{1-x}$ alloys can be grown on Ag(111) with a $(3 \times \sqrt{3})$ structure containing four atoms per unit cell (2/3 ML total coverage) \cite{Fuhr2024}. Remarkably, the crystallographic structure remains approximately constant over a wide range of $x$, spanning from Bi-rich to Sb-rich stoichiometries. This allowed us to investigate the effect of inversion symmetry breaking induced by substituting a heavier atom (Bi) with a lighter isoelectronic one (Sb) within a fixed crystallographic framework.

Here, we report a detailed study of the electronic structure of a Bi-rich Bi$_x$Sb$_{1-x}$ layer grown on Ag(111), combining angle-resolved photoemission spectroscopy (ARPES), X-ray photoelectron spectroscopy (XPS), and density functional theory (DFT) calculations. Our results show that replacing Bi atoms with lighter Sb ones in the $(3 \times \sqrt{3})$ structure does not induce significant changes in the overall electronic structure. Nevertheless, Sb incorporation enhances spin splitting along the surface normal by introducing in-plane asymmetry in the electronic potential. These findings provide new insight into the relationship between alloy composition, interfacial structure, and spin-orbit coupling in two-dimensional materials.

\section{Experimental and computational details}
\label{expdetails}

The experiments were performed at the VUV-photoemission beamline of the Elettra synchrotron (Trieste, Italy).

\subsection{Sample Preparation.}
The Ag(111) substrate (purchased from XXX) was cleaned following standard procedures consisting of repeated cycles of 1~keV Ar$^+$ sputtering and annealing at 450$^{\circ}$C, until a sharp $(1\times1)$ LEED pattern was observed. The base pressure during preparation was maintained below $5\times10^{-10}$~mbar.

In this study, the thickness of the Bi$_x$Sb$_{1-x}$ films is expressed in monolayers (ML), where 1~ML is defined as the atomic density of an unreconstructed Ag(111) surface, corresponding to $1.39 \times 10^{15}$~atoms/cm$^{2}$. Bi$_x$Sb$_{1-x}$ monolayers with controlled stoichiometries were prepared by simultaneous evaporation of ultrapure bismuth and antimony (Sigma-Aldrich) onto the Ag(111) substrate held at room temperature (RT), using a custom-made double Knudsen cell. The evaporation rates of the two sources were independently calibrated using a quartz crystal microbalance and by referencing the $(\sqrt{3}\times\sqrt{3})R30^{\circ}$ LEED patterns of the corresponding 1/3~ML surface alloys. Typical deposition rates ranged from $0.018$ to $0.055$~ML/min, allowing precise control of the Bi/Sb ratio.

After deposition, the sample quality and stoichiometry were verified by LEED and X-ray photoelectron spectroscopy (XPS). LEED patterns confirmed the formation of well-ordered $(3\times\sqrt{3})$ surface structures, while XPS measurements were used to determine the relative Bi/Sb composition. The base preassure in the preparation chamber was  $\le 3\times 10^{-10}$~Torr. 

\subsection{ARPES measurements}
Angle-resolved photoemission spectroscopy (ARPES) measurements were performed using a Scienta R4000 electron analyzer and excitation energies in the range between  60 and 120 eV with linearly polarized light, while keeping the sample at 18 K. The electron spectrometer was placed at \textbf{35$^{\circ}$ ?} with reference to the direction of the incoming photon beam. The energy resolution was about \textbf{20 meV ?}, and the angular resolution better than \textbf{0.3$^{\circ}$ ?}. The base pressure of the analytic was $\le 1.0\times 10^{-10}$~Torr. 

\subsection{ Camputational details} 
The density functional theory (DFT) calculations were performed using the Quantum ESPRESSO package,\cite{Q-E} which is a plane-wave implementation with pseudopotentials.
In all calculations, we used the PBE\cite{PBE} exchange correlation potential with Grimme-D2\cite{Grimme-D2} van der Waals interactions. We used fully relativistic PAW pseudopotentials, which include spin-orbit interaction, and a wave function/density cutoff of 45/450 Ry. Brillouin integrations were conducted using a uniform $\Gamma$-centered $k$-point mesh of $30\times 30\times 30$ for bulk Ag. With these parameters, we obtained an optimized lattice parameter for Ag bulk of 4.135~\AA, which we used in the subsequent of the calculations. The convergence threshold for electronic self-consistency was set to 10$^{-7}$ a.u., while the convergence threshold on total energy for ionic minimization was set to 10$^{-5}$ a.u.

For the surface calculations, we used the slab method with nine pure Ag layers and the top Sb-Bi layer. We used a vacuum layer larger than 15~\AA\ between replicas along the $z$ direction to minimize interactions. In all the calculations, we fixed the two lower layers while all other atoms are allowed to relax. Being consistent with the bulk calculations, for the $(3 \times \sqrt{3})$ surface unit cell, we used for Brillouin integrations a uniform $\Gamma$-centered $k$-point mesh of $10\times 17\times 1$.

\section{\label{Resu}Results and Discussion}

\subsection{The Bi$_x$Sb$_{1-x}$/Ag(111)-$(3 \times \sqrt{3})$ surface}

\begin{figure}
\includegraphics[width=1\columnwidth] {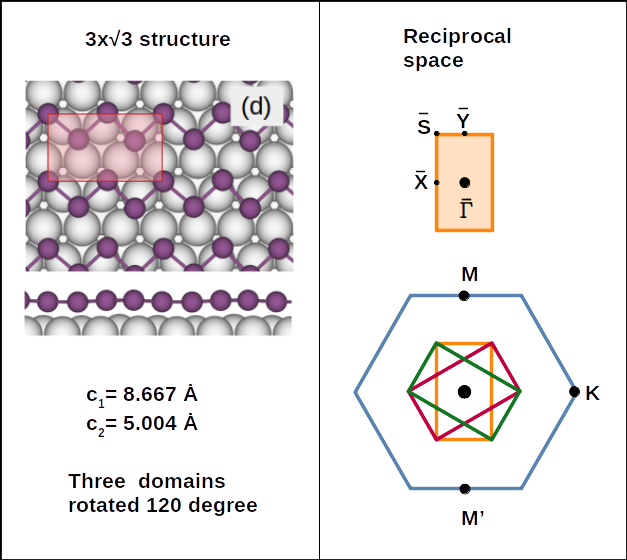}
 \caption{Left: Schematic representation of the $(3\times\sqrt{3})$ surface structure. Owing to the absence of chemical ordering, the purple circles represent either Bi or Sb atoms. The total (Bi+Sb) coverage is 2/3 ML. Right: Corresponding reciprocal space. The $\bar{\Gamma}$–$\bar{Y}$, $\bar{\Gamma}$–$\bar{X}$, and $\bar{\Gamma}$–$\bar{S}$ distances are 0.625, 0.36, and 0.72~\AA$^{-1}$, respectively.}. 
  \label{fig:Model}
\end{figure}

Previously, we demonstrated that simultaneous evaporation of Bi and Sb onto a Ag(111) surface yields single-layer Bi$_x$Sb$_{1-x}$ films with a rectangular $(3\times\sqrt{3})$ structure and variable stoichiometry.\cite{Fuhr2024} For a total coverage (Bi+Sb) of 2/3 ML, scanning tunneling microscopy (STM) images revealed high-quality surfaces consisting of large $(3\times\sqrt{3})$ domains, albeit with evident chemical disorder. A schematic model of the corresponding atomic structure is shown in Fig.~\ref{fig:Model}, together with the associated reciprocal lattice. This $(3\times\sqrt{3})$ structure, containing four atoms per unit cell, is similar to those formed by pure Sb and Bi overlayers at the same coverage and room temperature. Structurally, the main difference between the Sb and Bi overlayers is that Bi forms an incommensurate rectangular ($p\times\sqrt{3}$) structure.\cite{Pussi2011} Consequently, Sb incorporation stabilizes a commensurate surface reconstruction.

\begin{figure*}[ht]
  \centering
\includegraphics[width=0.9\textwidth]{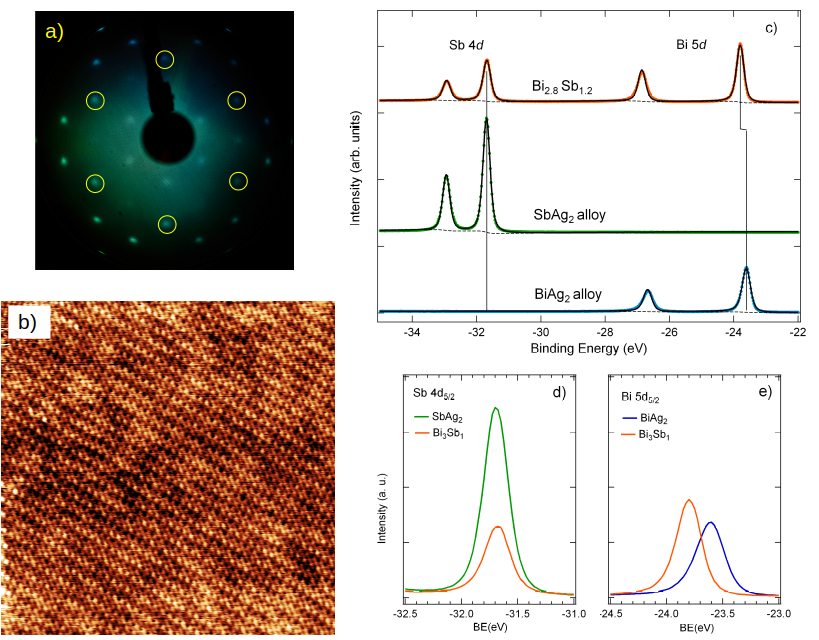}
 \caption{(a) LEED pattern of the Bi-rich Bi$_x$Sb$_{1-x}$ single-layer surface acquired at an incident electron energy of 60 eV. Yellow circles highlight diffraction spots coinciding with the $(\sqrt{3}\times\sqrt{3})$ reconstruction. (b) STM image of a Bi-rich surface. Size: $25 \times 25$ nm$^{2}$. Tunneling conditions: +0.2V/1.0 nA. (c) Experimental photoemission spectra (colored dots) recorded at a photon energy of 70 eV for the two pure surface alloys and the Bi-rich surface, along with the corresponding fits (black lines) and background (dashed black lines). The spectra are normalized to the intensity of the background at 22 eV. (d) and (e) Details of the experimental spectra in the regions of the Sb~$4d$ and Bi~$5d$ core levels, respectively.
%At this photon energy, the photoemission cross section of Sb 4$d$ is larger than that of Bi 5$d$ by a factor of approximately 1.74.
}
\label{fig:LEEDXPS}
\end{figure*}

Following this procedure, we grew a Bi-rich Bi$_x$Sb$_{1-x}$ overlayer with a $(3\times\sqrt{3})$ structure. A representative LEED pattern is shown in Fig.~\ref{fig:LEEDXPS}(a). The LEED image displays sharp diffraction spots characteristic of the $(3\times\sqrt{3})$ structure, together with a low background, indicating a high degree of long-range order. It is worth noting that a subset of the spots highlighted in yellow, associated with the $(3\times\sqrt{3})$ superstructure, coincides with those of the $(\sqrt{3}\times\sqrt{3})$ surface alloys. Fig.~\ref{fig:LEEDXPS}(b) shows a typical STM image of the Bi-rich Bi$_x$Sb$_{1-x}$/Ag(111)-$(3\times\sqrt{3})$ surface. The core-level spectra of the Bi-rich sample are compared with those of the SbAg$_2$ and BiAg$_2$-$(\sqrt{3}\times\sqrt{3})R30^{\circ}$ surface alloys in Figs.~\ref{fig:LEEDXPS}(c)-(e).

The calculated differential cross-section ratio, $\sigma(\text{Sb }4d)/\sigma(\text{Bi }5d)$, is 1.74. The relative areas of the Bi~$5d$ and Sb~$4d$ core-level peaks, weighted by their respective photoemission cross sections, yield a composition of Bi$_{2.8}$Sb$_{1.2}$. The fitting analysis of the Bi~$5d_{5/2}$ and Sb~$4d_{5/2}$ core-level spectra reveals narrow single components, consistent with the formation of a mixed overlayer without detectable Ag incorporation~\cite{Pussi2011,Fuhr2024}. The Bi~$5d$ binding energy is shifted by $+0.2$~eV relative to that in the BiAg$_2$ surface alloy, while no shift is detected for the Sb~$4d$ level.

\subsection{ARPES Data Analysis}

The electronic structure of the prepared $(3\times\sqrt{3})$–Bi$_{1-x}$Sb$_x$/Ag(111) surfaces was investigated by measuring binding energy (BE) versus $k_{\parallel}$ photoemission spectra along the $[1\bar{1}0]$ ($\Gamma$–$K$), $[11\bar{2}]$ ($\Gamma$–$M$), and $[\bar{1}\bar{1}2]$ ($\Gamma$–$M$') directions of Ag(111), using photon energies of 60 and 120 eV. Here, $\Gamma$, $M$ ($M$'), and $K$ denote high-symmetry points of the Ag(111) surface Brillouin zone, corresponding to normal emission and parallel momenta of 1.24 and 1.45~\AA$^{-1}$, respectively. Additionally, three-dimensional ARPES data ($E_\mathrm{B}$, $\theta$, $\phi$) were collected for both samples at a photon energy of 60 eV.

%%%%%%%%%%%BULK

\begin{figure*}[ht]
  \centering
\includegraphics[width=0.8\textwidth]{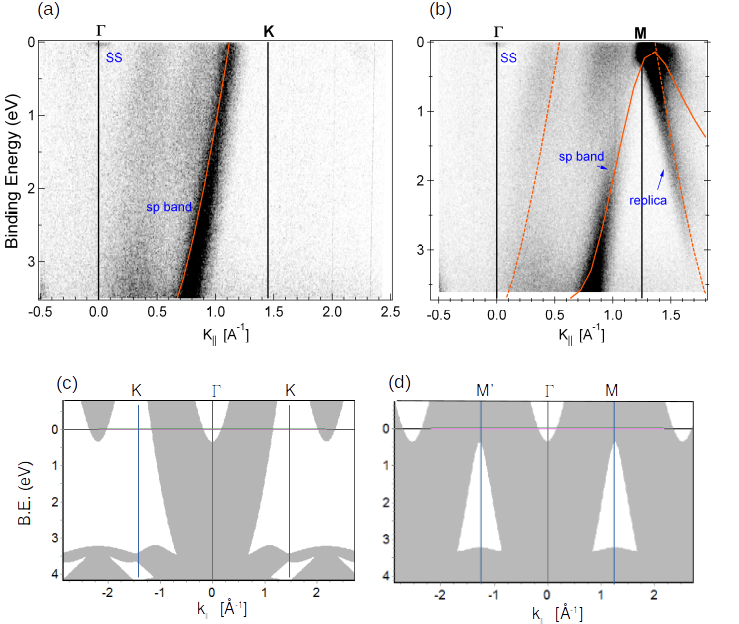}
 \caption{(a) BE vs $k_{\parallel}$ photoemission spectra acquired along the $[1\bar{1}0]$  direction ($\Gamma$–$K$) at a photon energy of 120 eV. The calculated $sp$ band is shown as a orange solid line, and its Umklapp replicas as orange dashed lines. (b) Same as (a), but along the $[11\bar{2}]$ direction ($\Gamma$–$M$). In (a) and (b), the photoemission intensity is displayed in reversed grayscale (black corresponds to high intensity). (c) and (d) Projections of the bulk Ag states onto the (111) surface plane along the $\Gamma$–$K$ and $\Gamma$–$M$ directions, respectively. }
\label{fig:Agbulk}
\end{figure*}

To identify bulk-derived Ag features, spectra from clean Ag(111), acquired under identical conditions, were used as a reference. In particular, BE versus $k_{\parallel}$ maps measured with a photon energy of 120 eV along the $\Gamma$–$K$ and $\Gamma$–$M$ directions are shown in Figs.~\ref{fig:Agbulk}(a) and (b), respectively. The intense features observed in both maps are attributed to contributions from the bulk sp band, whose calculated dispersion \citep{VJoco} is overlaid as orange solid lines. Electrons originating from the bulk sp band can undergo Umklapp processes on their way out of the solid, exchanging momentum with the topmost Ag(111) layers and giving rise to replica bands. A particularly intense replica is observed in Fig.~\ref{fig:Agbulk}(b), where it can be reproduced (orange dashed lines) by shifting the $sp$ band along the $[11\bar{2}]$ direction by 2.51~\AA$^{-1}$, corresponding to the magnitude of the surface reciprocal lattice vector of the hexagonal unit cell.

In addition, the Shockley surface state of the bulk-terminated substrate is visible in both BE versus $k_{\parallel}$ maps shown in Fig.~\ref{fig:Agbulk}, appearing as a characteristic oval feature around the $\Gamma$ point at the Fermi level, and indicating the formation of a well-ordered Ag(111) surface. This feature becomes significantly more intense at a photon energy of 60 eV, as can be seen in Figures~S2 and S3 of Supporting Information.

To aid the analysis of these ARPES maps, the bulk band gaps are identified using the projected bulk band structure. Figs.~\ref{fig:Agbulk}(c) and (d) display the projection of the bulk Ag states onto the (111) surface plane along the $\Gamma$–$K$ and $\Gamma$–$M$ directions, respectively. The energy gap around the $\Gamma$ point, as well as those near the $M$ and $K$ points, are clearly visible.

%%%%%%%%%%%%%%%%%%%%%%%%%%%%%%%%%%%%%%3xR3%%%%%%%%%%%%%%%%%%%%%%%%%%%%%%%%%%% GK

\begin{figure*}[ht]
  \centering
  \includegraphics[width=0.7\textwidth]{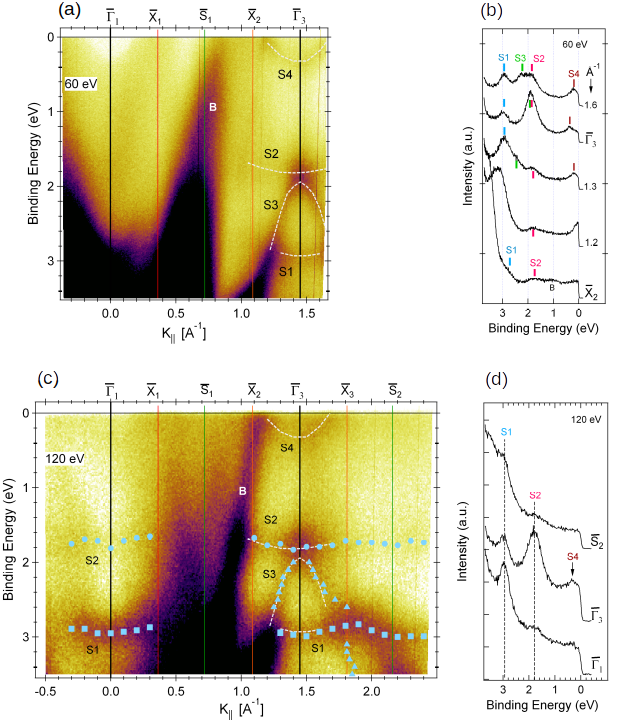}
 \caption{(a) and (c) BE vs $k_{\parallel}$ photoemission spectra of the Bi-rich $(3\times\sqrt{3})$ surface acquired along the $[1\bar{1}0]$ direction ($\Gamma$–$K$) at photon energies of 60 and 120 eV, respectively. The $\bar{\Gamma}_3$ point of the reconstruction coincides with $K$. Photoemission intensity is represented in a reversed yellow-purple coler scale, i.e., darker corresponds to higher photoelectron intensity.
(b) and (d) Representative vertical line profiles extracted from the maps in (a) and in (c), respectively, at selected $k{\parallel}$ values. }
  \label{fig:TeoExpK}
\end{figure*}

\begin{figure}
\includegraphics[width=1.0\columnwidth] {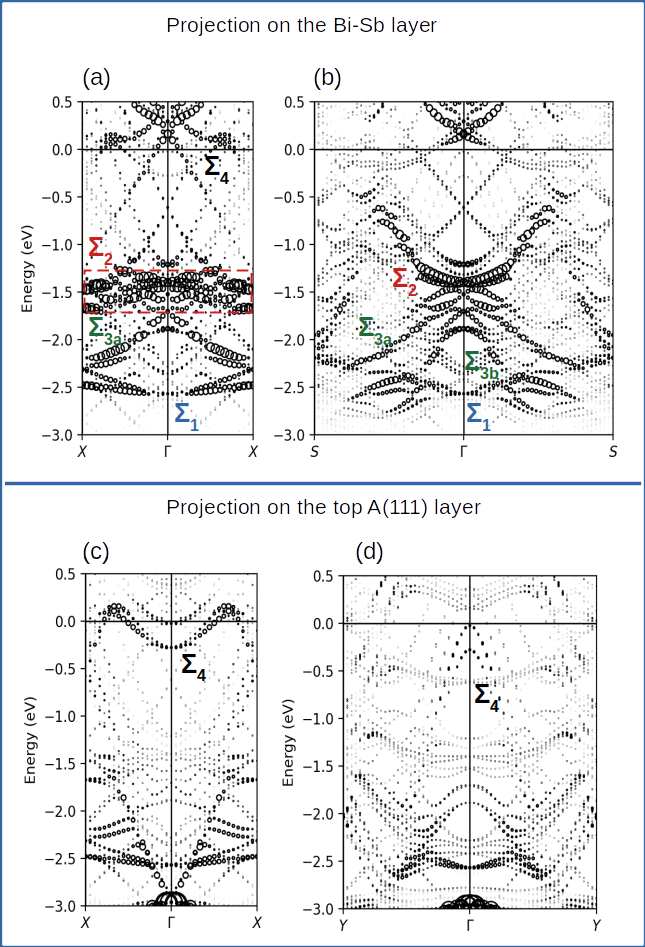}
\caption{(a),(b) Calculated SOC band structure for the ordered Bi$_3$Sb$_1$/Ag(111) surface along $\bar{\Gamma}$–$\bar{X}$ and $\bar{\Gamma}$–$\bar{S}$, respectively, projected onto the Bi-Sb overlayer. (c),(d) Corresponding band structures along $\bar{\Gamma}$–$\bar{X}$ and $\bar{\Gamma}$–$\bar{Y}$, projected onto the top Ag(111) layer. Symbol size is proportional to the projection weight in all panels.}
  \label{fig:CalculosGK}
\end{figure}

We now focus on the $(3\times\sqrt{3})$ surface with an overlayer with a Bi-rich composition. Figures~\ref{fig:TeoExpK}(a) and (c) show BE–$k_{\parallel}$ photoemission maps measured along the $\Gamma$–$K$ direction at photon energies of 60 and 120 eV, respectively. In the following, the photoemission intensity is displayed using a yellow–purple color scale (purple denotes the highest intensity).

Owing to the threefold rotational symmetry of the $(3\times\sqrt{3})$ reconstruction, three equivalent rectangular domains coexist on the surface, resulting in a complex reciprocal-space pattern (Fig.~S1 of the Supporting Material). From the viewpoint of the $(3\times\sqrt{3})$ reconstruction, the $\Gamma$ and $K$ points of the Ag(111) clean surface are equivalent and both coincide with the $\bar{\Gamma}$ point of all three domains, overbars denoting high-symmetry points of the $(3\times\sqrt{3})$ structure. Along the $\Gamma$–$K$ direction, the Brillouin zones of these domains are intersected along different high-symmetry directions. The aligned domain (orange in Fig.~\ref{fig:Model}) is probed along $\bar{\Gamma}$–$\bar{X}$, and the inclined domains (green and red in Fig.~\ref{fig:Model}) are probed along $\bar{\Gamma}$–$\bar{S}$. In Figures~\ref{fig:TeoExpK}(a) and (c) we marked the repeated $\bar{X}$ and $\bar{S}$ corresponding points appearing along the $\Gamma$-$K$ direction.

At 60 eV, the BE–$k_{\parallel}$ map clearly resolves the bulk gap at $\Gamma$, with no detectable Shockley surface state. Instead, bands S1–S4 emerge within the bulk gap around $K$ (coinciding with $\bar{\Gamma}_3$) and are assigned below to surface states of the $(3\times\sqrt{3})$ reconstruction. Figure~\ref{fig:TeoExpK}(b) shows representative vertical line profiles from the 60 eV map at different $k{\parallel}$. The S1–S4 bands appear as broad, weak features , yet can be tracked across the $\bar{\Gamma}_3$–$\bar{X}_2$ range, allowing their dispersion to be extracted. The resulting dispersion is overlaid as yellow dashed lines on the BE–$k_{\parallel}$ maps. As discussed below, the pronounced broadening of these bands primarily reflects the lack of long-range chemical order in the Bi$_x$Sb$_{1-x}$ overlayer. \cite{Gierz2011}

At 120 eV, the accessible $k_{\parallel}$ range is significantly extended, while the bulk contribution around $\Gamma$ is strongly suppressed, mainly due to the Cooper minimum of the Ag~4$d$ photoemission cross section, thereby enhancing the visibility of surface states. Figure~\ref{fig:TeoExpK}(d) shows representative vertical line profiles extracted from the 120 eV map at relevant $k_{\parallel}$ values. A corresponding analysis yields the marker curves shown in Fig.~\ref{fig:TeoExpK}(c). Notably, the S1 and S2 bands observed around $\bar{\Gamma}_3$ are replicated at $\bar{\Gamma}_1$, consistent with the symmetry of the $(3\times\sqrt{3})$ reconstruction. The excellent agreement between the 60 eV (dashed lines) and 120 eV (markers) demonstrates that the S1–S4 bands are photon-energy independent, as expected for surface states.

To further elucidate the electronic band structure of the $(3\times\sqrt{3})$ overlayer, we carried out DFT calculations including spin-orbit coupling for a Bi-rich Bi$_3$Sb$_1$ surface. Figures~\ref{fig:CalculosGK}(a) and (b) display the calculated dispersions along the $\bar{\Gamma}$–$\bar{X}$ and $\bar{\Gamma}$–$\bar{S}$ directions projected onto the Bi-Sb overlayer. The calculations yield multiple bands with significant overlayer weight, with some regions where they cluster into groups of surface states. This is expected given the reduced symmetry of the system, arising from a rectangular overlayer supported on a hexagonal substrate, as well as from the significant interaction between the overlayer $p_z$ orbitals and the Ag bulk states.

In both directions, a high density of bands with large overlayer weight is found at the $\bar{\Gamma}$ point approximately 1.5~eV below the Fermi level (grouped under the label $\Sigma_2$). These correlate — aside from a compression of the energy scale present in the DFT calculations — with the large feature observed experimentally at the $K$ point in the range 1.6–2.0~eV below the Fermi level. From this point, the calculated bands with flat (upward) dispersion along $\bar{\Gamma}$–$\bar{X}$ ($\bar{\Gamma}$–$\bar{S}$) contribute to the experimental feature $S_2$. The experimental band $S_1$ can be identified with the low-dispersion calculated bands ($\Sigma_1$) appearing in both directions, with a minimum around 2.6~eV below the Fermi level. Finally, the observed $S_3$ band can be identified with the multiply calculated bands ($\Sigma_{3a}$ and $\Sigma_{3b}$) exhibiting downward dispersion from the $\bar{\Gamma}$ point in both directions.

In contrast, the experimentally observed $S_4$ band appears along $\bar{\Gamma}$–$\bar{X}$ (as $\Sigma_4$) with very low overlayer weight, yet with a high projection onto the first Ag(111) layer (see Fig.~\ref{fig:CalculosGK}(c)), indicating that $S_4$ has an interface character. Complementarily, Fig.~\ref{fig:CalculosGK}(d) shows the calculated band structure projected onto the top substrate layer along the $\bar{\Gamma}$–$\bar{Y}$ direction, where two downward-dispersing bands centered at the $\bar{\Gamma}$ point appear with low surface weight  (indicating strong hybridization with bulk states ) and together suggest that $\Sigma_4$ has a saddle-point shape. We return to this point below.

% Along $\bar{\Gamma}$–$\bar{X}$ [Fig.~\ref{fig:TeoExpK}(e)], the calculations predict two bands at $\sim$2.5 eV: one nearly nondispersive and another exhibiting a weak upward dispersion. These features reproduce the experimental S1 band, aside from a slight compression of the energy scale. Nonaligned domains also contribute significantly to the S1 feature, as indicated by Fig.~\ref{fig:TeoExpK}(f). In particular, a wing-like band centered at $\bar{\Gamma}$, with a minimum near 2.5 eV, is predicted, accounting for the S1 feature, especially in the 120 eV map.

% The calculated $\bar{\Gamma}$–$\bar{X}$ dispersion further reveals a cluster of surface states around $\bar{\Gamma}$ at binding energies of $\sim$1.4 eV. Together, these states form a broad feature with weak downward dispersion. In contrast, the $\bar{\Gamma}$–$\bar{S}$ dispersion [Fig.~\ref{fig:TeoExpK}(f)] exhibits a band with a minimum near 1.4 eV and a pronounced upward dispersion, accompanied by a strong reduction in spectral weight with increasing $k_{\parallel}$. The combined contributions of these calculated features account for the experimental S2 band, including its broad character and weak overall dispersion.

% The S3 band is also well reproduced, as both $\bar{\Gamma}$–$\bar{X}$ and $\bar{\Gamma}$–$\bar{S}$ cuts exhibit downward-dispersing features in good agreement with experiment.

%&&&&&&&&&&&&&&&&&&&&&&&&&&&&&&&&&&&&&&&&&&&&   GM

\begin{figure}
\includegraphics[width=1.0\columnwidth] {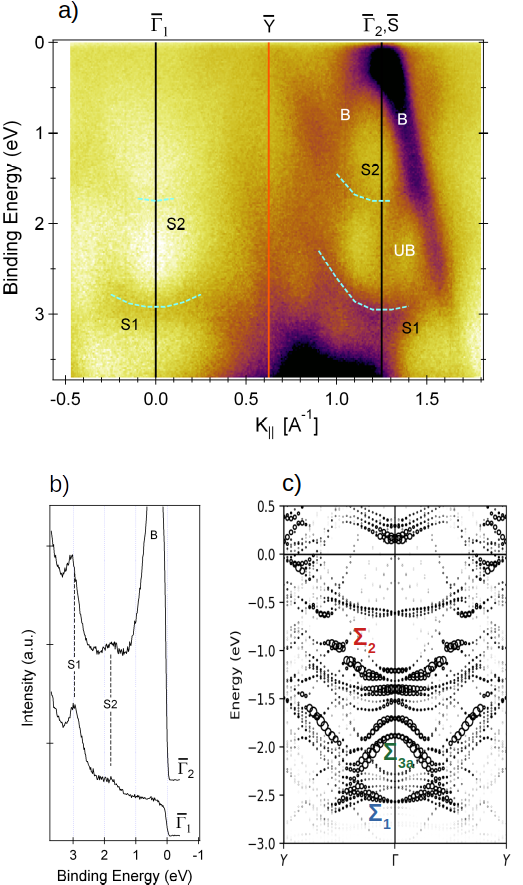}
\caption{(a) BE vs $k_{\parallel}$ photoemission spectra of the Bi-rich Bi$_x$Sb$_{(1-x)}$/Ag(111) surface acquired with a photon energy of 120 eV along $[11\bar{2}]$ ($\Gamma$–$M$ direction). The position of the $\bar{Y}$ point of the aligned $(3\times\sqrt{3})$ domain (orange in Fig.~\ref{fig:Model}) is indicated. See text for details.
(b) Vertical intensity profiles extracted from the  map in (a).
(c) Calculated band structure of the ordered Bi$_3$Sb$_1$/Ag(111) surface along the $\bar{\Gamma}$–$\bar{Y}$ direction, projected onto the overlayer and computed with SOC. The circle size is proportional to the projection weight. }
  \label{fig:TeoExpM}
\end{figure}

\begin{figure*}[ht]
  \centering
  \includegraphics[width=0.8\textwidth]{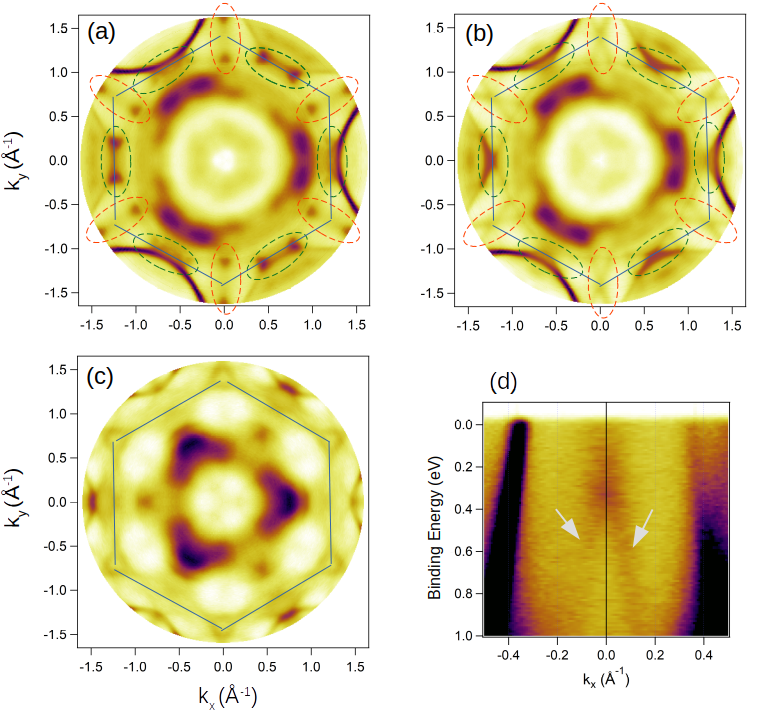}
 \caption{ (a)–(c) Constant-energy maps from the Bi-rich sample measured at $h\nu = 60$ eV  for binding energies $E_B = 0$ (Fermi level), 0.33, and 2.17 eV. The maps were obtained from a three-dimensional set of experimental data collected over a  $100^{\circ}$ azimuthal range and symmetrized to the full map. The $k_x$ and $k_y$  axes are oriented along the $\Gamma$–$M$ and $\Gamma$–$K$ directions of the Ag(111)  surface Brillouin zone, respectively. The surface Brillouin zone of the Ag(111)  substrate is indicated in green. (d) BE vs. $k_{\parallel}$ cut at $k_y = 1.45$ Å$^{-1}$, corresponding to the $\bar{\Gamma}$–$\bar{Y}$ direction. The arrows highlight a downward-dispersing band attributed to S4. }
  \label{fig:FS}
\end{figure*}

Figure~\ref{fig:TeoExpM}(a) shows the BE–$k_{\parallel}$ map measured at  120 eV along the substrate $\Gamma$–$M$ direction; the ARPES 
data obtained at 60 eV are shown in Fig.~S2 of the Supporting Material.  Given the close similarity between these data and those collected along the  $\Gamma$–$M'$ direction (Fig.~S3 of the Supporting Material),  the following analysis focuses on the former case, as similar conclusions can be drawn from both data sets.

In analyzing the ARPES data measured along $\Gamma$–$M$, it is important  to note that, for the reconstructed surface, the substrate $\Gamma$ and $M$ points are not equivalent (see Fig.~S1 of the Supporting Material).
The $\Gamma$ point coincides with the $\bar{\Gamma}$ point of all three domains, whereas the substrate $M$ point maps onto a $\bar{\Gamma}$ point of the aligned domain but onto $\bar{S}$ points of the two inclined domains. Consequently, the substrate $\Gamma$ and $M$ points are equivalent for the aligned domain but inequivalent for the inclined domains. Accordingly, the $\Gamma$–$M$ cut corresponds to the $\bar{\Gamma}$–$\bar{Y}$ high-symmetry direction for the aligned domain, whereas for the inclined domains it does not follow a high-symmetry path. The dashed lines overlaid on the experimental ARPES map were obtained from an analysis of vertical line profiles, including those shown in Fig.~\ref{fig:TeoExpM}(b), following the procedure described above.

Remarkably, the region around normal emission ($\bar{\Gamma}$ point) is largely free of Ag bulk-state contributions. In this region, the map reveals two broad bands with minima at binding energies of approximately 2.9 and 1.75~eV [Fig.~\ref{fig:TeoExpM}(b)], assigned to the $S_1$ and $S_2$ bands discussed above. These bands are also observed at the $M$ point ($\bar{\Gamma}_2$), reflecting the reciprocal-space periodicity of the aligned domain. The Ag bulk gap around the $M$ point lies within the triangular region defined by the bulk bands indicated in Fig.~\ref{fig:Agbulk}(b). Within this gap, the steeply upward-dispersing feature appearing between the $S_1$ and $S_2$ bands is attributed to a bulk-related band arising from umklapp processes associated with the $(3\times\sqrt{3})$ reconstruction. Comparison with the calculated Bi-Sb projected bands along $\bar{\Gamma}$–$\bar{Y}$ [Fig.~\ref{fig:TeoExpM}(c)] shows that $S_1$ and $S_2$ are well reproduced around the $\bar{\Gamma}$ and $M$ points by the calculated $\Sigma_1$ and $\Sigma_2$ bands, respectively. The latter predicts a pronounced upward dispersion for $S_2$, reaching the Fermi level near the $\bar{Y}$ point. However, this prediction is difficult to corroborate experimentally due to the low spectral weight of the $S_2$ band and the dominant contribution of Ag bulk states.

%The Ag bulk gap around the $M$ point lies within the triangular region defined by the indicated bulk bands, as shown in Fig. \ref{fig:Agbulk}(b). Within this gap, the two pronounced upward-dispersing bands are correspond to the S1 and S2, while an additional steeply upward-dispersing feature appears inside the gap and is attributed to a bulk-related band arising from umklapp processes associated with the $(3\times\sqrt{3})$ reconstruction.

% The BE–$k_{\parallel}$ map measured along the substrate $\Gamma$–$M'$ direction (see Fig.~S7 of the Supporting Material) is essentially equivalent to the $\Gamma$–$M$ map discussed above and therefore does not provide additional information.

%=========================Fermi Surface

Figure~\ref{fig:FS} shows constant-energy maps from the Bi-rich sample at photon energy $h\nu = 60$~eV for selected binding energies. The Fermi surface map in panel~(a) is dominated by contributions from the bulk $sp$ band, as confirmed by comparison with the calculated Fermi surface shown in Fig.~S4(b) of the Supplemental Material. As discussed above in connection with the BE vs $k_{\parallel}$ maps in Figs.~\ref{fig:TeoExpK} and~\ref{fig:TeoExpM}, the only band of the prepared surface with significant spectral weight at the Fermi level is the interfacial $S_4$ band, which gives rise to the two prominent features near the $K$ points (indicated by red dashed ovals). As discussed in connection with Figs.~\ref{fig:CalculosGK}(c) and~\ref{fig:CalculosGK}(d), calculations suggest that $S_4$ has a saddle-point shape. This is confirmed by the binding energy vs $k_{\parallel}$ cut at $k_y = 1.45$~Å$^{-1}$ shown in Fig.~\ref{fig:FS}(d), where a downward-dispersing feature can be identified with the $\Sigma_4$ band along $\bar{\Gamma}$–$\bar{Y}$. 
 
The surface origin of the interfacial $S_4$ state is further supported by the observation of replicas imposed by the $(3\times\sqrt{3})$ reconstruction. A second set of intensity maxima at the $M$ and $M'$ points of the Fermi surface (indicated by blue dashed ovals) is consistent with such replication (see (Fig.~S1 of the Supplemental Material).  This correspondence is further corroborated by the $E_\mathrm{B}$–$k_{\parallel}$ maps along the $K$–$M$–$K$ and $K$–$M'$–$K$ directions extracted from the three-dimensional ARPES data set, which reveal a parabolic band closely resembling the $S_4$ state (Fig.~S4(c)-S4(e) of the Supplemental Material). Replicas expected at smaller $k$ values are not observed, which we attribute to their reduced spectral weight.

Figure~\ref{fig:FS}(b) shows a constant-energy map taken at $E_\mathrm{B} = 0.33$ eV, corresponding to the saddle point of the $S_4$ band. At the $K$ points, clear X-shaped features are visible, which are replicated at the $M$ and $M'$ points following the same pattern as the $S_4$-derived features discussed above, and can all be attributed to the $S_4$ band.

Finally, the constant-energy map in panel~(c) ($E_\mathrm{B} = 2.17$~eV) shows closed contours at the $K$ points associated with the downward-dispersing $S_3$ band. Notably, analogous features are absent at the $M$ and $M'$ points,  an asymmetry we attribute to the inequivalence of the $M$ and $K$ points with respect to the $(3\times\sqrt{3})$ superstructure. While all three domains contribute to the features near $K$, only a single domain contributes near $M$; combined with the strong bulk-state spectral weight in the vicinity of $M$, this provides a natural explanation for the absence of replica features at that point.

%=========================Sb-Rich

%We now turn to the stoichiometrically balanced Bi$_x$Sb$_{1-x}$ sample. The ARPES data obtained from this sample (Fig.~S5 of the Supplemental Material) reveal a band structure closely resembling that of the Bi-rich case, as expected from the similar electronic properties of Bi and Sb. This experimental similarity is supported by calculations comparing the electronic band structures of ordered $(3 \times \sqrt{3})$–Bi$_3$Sb$_1$ and $(3 \times \sqrt{3})$–Bi$_2$Sb$_2$ surfaces (Fig.~S6 of the Supplemental Material), which show small but discernible shifts in band positions. Although modest, these stoichiometry-dependent shifts likely contribute to the observed band broadening, given that the surface is not expected to be chemically homogeneous.
% Therefore, this intrinsic inhomogeneity is identified as the primary origin of the pronounced broadening of the experimentally observed bands.

In summary, three surface-state bands ($S_1$–$S_3$) with strong weight on the Bi-Sb overlayer are experimentally identified, along with an interfacial band ($S_4$) with strong weight on the top substrate layer. The latter exhibits a saddle point and is responsible for prominent features in the Fermi surface. The $S_2$ band may additionally reach the Fermi level near the $\bar{Y}$ point, albeit with low spectral weight. The observed band broadening arises from several factors: the reduced symmetry of the rectangular overlayer on the hexagonal substrate, the significant coupling of Bi-Sb $p_z$ orbitals with substrate bands, and the lack of long-range chemical order.

% Finally, the overall agreement supports the proposed structural model of a rectangular four-atom overlayer unit cell.
 
\subsection{Theoretical Analysis}

%%%%%%%%%%%%%%%%%%%%%%%%%%%%%%%%%%%%%%%%%%
% CALCULATIONS

\begin{figure}[ht]
\includegraphics[width=0.6 \columnwidth]{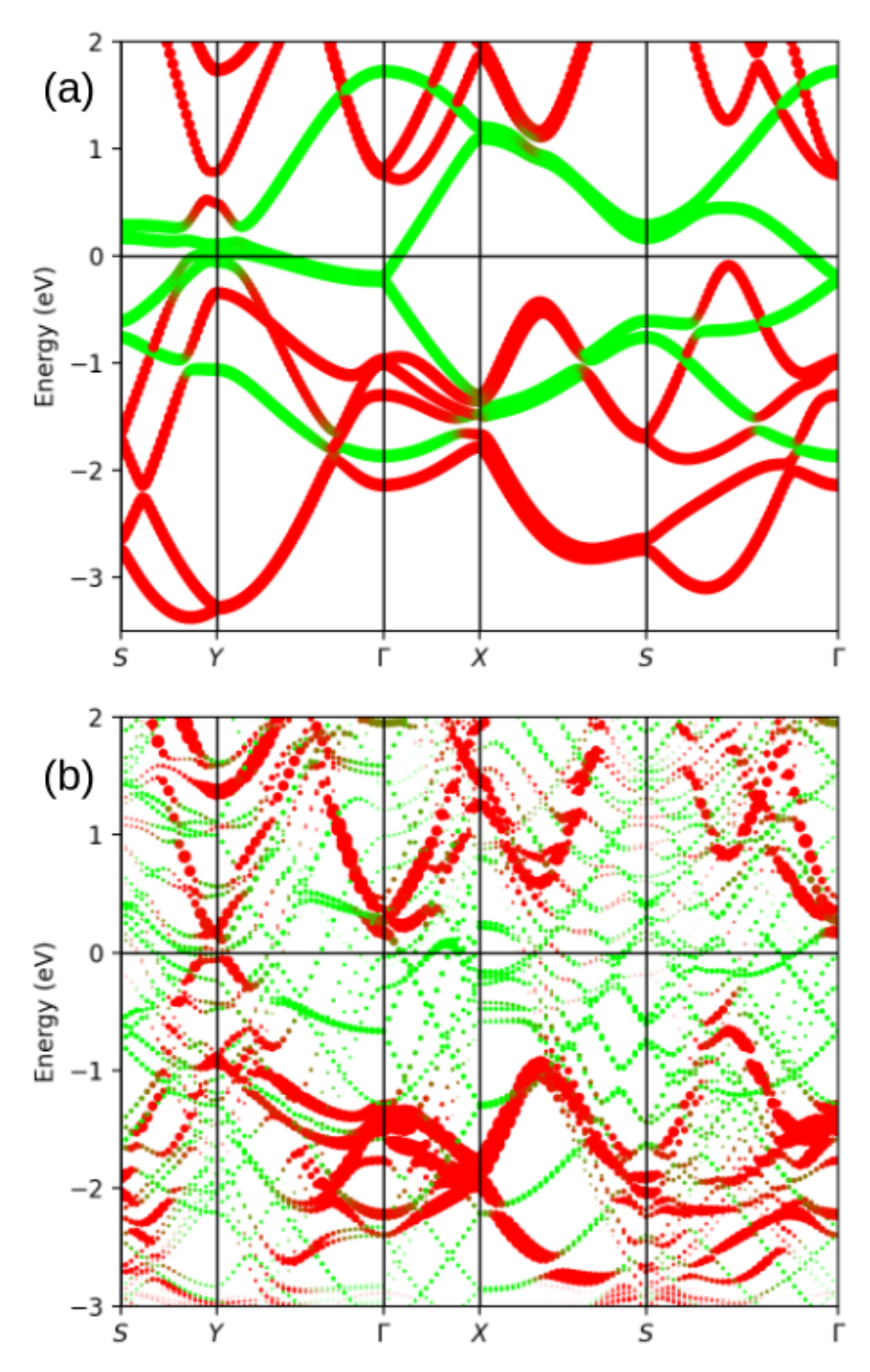}
\caption{Calculated band structures without spin–orbit coupling. (a) Free-standing Bi$_3$Sb$_1$ layer adopting the atomic geometry of the Ag(111)-supported system. (b) Bi$_3$Sb$_1$ layer adsorbed on Ag(111), with the symbol size proportional to the corresponding state weight in the Bi–Sb layer. In both panels, the color scale indicates the orbital character, with red and green corresponding to $p_{xy}$ and $p_{z}$, respectively.}
  \label{fig:orbital}
\end{figure}

\begin{figure*}[ht]
  \centering
\includegraphics[width=1.0\textwidth]{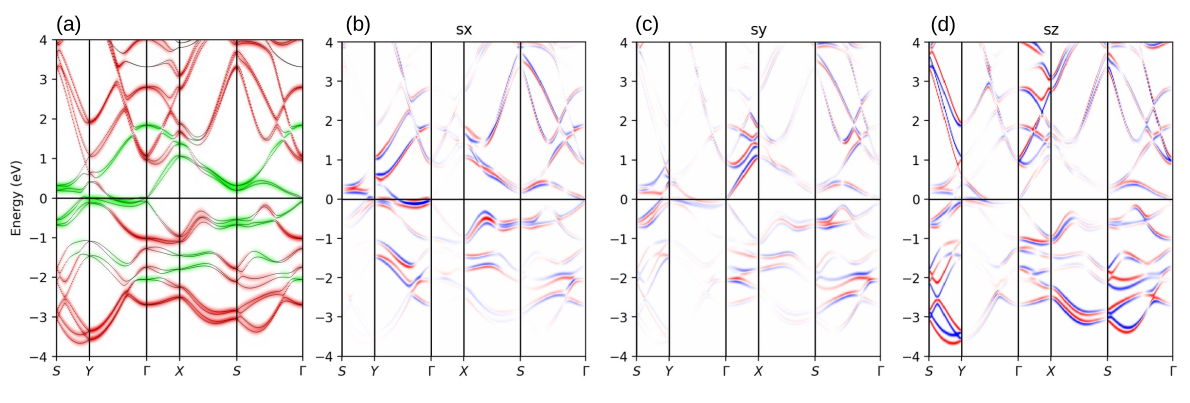}
 \caption{ (a) $k$-resolved orbital projected density of states for a free-standing Bi$3$Sb$1$ layer adopting the atomic geometry of the Ag(111)-supported system, calculated including spin-orbit coupling. (b)–(d) $k$-resolved spin projected density of states, along the three Cartesian directions, with the $\hat{z}$ direction corresponding to the surface normal. The magnitude and sign of the spin polarization are represented on a blue-white-red scale, where blue, white, and red correspond to -20\%, 0, and +20\%, respectively.}
\label{fig:SPINunsup}
\end{figure*}

To investigate the substrate effect on the Bi-Sb overlayer electronic structure, we compare in Figure~\ref{fig:orbital}(a) and (b) the orbital projected bands of a freestanding and Ag(111) supported Bi-Sb layer, respectively. The calculation was performed without SOC and the projection onto Bi-Sb $p_{xy}$ and $p_{z}$ orbitals are highlighted in green and red, respectively. For the free-standing layer (Fig.~\ref{fig:orbital}(a)), the bands exhibit a clear separation between $p_{xy}$ and $p_{z}$ bands, with only minor mixing at band crossings. When supported (Fig.~\ref{fig:orbital}(b)), the predominant $p_z$ bands become strongly hybridized with Ag bulk states, whereas the $p_{xy}$ bands are significantly less affected, largely retaining their dispersion relative to the free-standing case and undergoing an approximately rigid shift toward lower energies. This substrate-induced orbital filtering effect has been previously reported for bismuthene on Ag(111) \cite{Sun2022} and on SiC(0001) \cite{Reis2017}. The inclusion of spin–orbit coupling (SOC) induces orbital mixing in some of the bands, as can be seen in Fig.~\ref{fig:SPINunsup}(a) for the unsupported Bi-Sb layer. Although this mixing can reduce the orbital filtering effect, surface states with $p_{xy}$ character nevertheless remain far less affected by the substrate than those with $p_z$ character. Therefore, the experimentally observed surface bands S1–S3 consist of surface states with predominant Bi-Sb $p_{xy}$ character.

Spin splitting is nearly absent in a pure Bi$_4$ layer adopting the atomic geometry of a $(3\times\sqrt{3})$–Bi$_4$/Ag(111) surface (Fig.~S7 of the Supporting Material). The inclusion of Sb atoms drastically change the spin splitting of the bands, as can be seen in Figures~\ref{fig:SPINunsup}(b)–(d) where we show the spin-resolved band structures of the freestanding Bi$_3$Sb$_1$ layer, with the spin polarization projected along the three Cartesian axes. We can see that the Bi$_3$Sb$_1$ layer exhibits important spin splitting with large spin polarization, both in-plane and out-of-plane.

The physics underlying Rashba-like spin splitting of surface-state bands 
indicates that breaking inversion symmetry along the surface normal gives 
rise to an in-plane spin polarization perpendicular to $k_{\parallel}$.\cite{Rashba1983,OAMPRL2011,OAMPRB2013}
In addition, in-plane asymmetries in the electrostatic potential can induce  spin splitting accompanied by a finite out-of-plane spin component $S_z$\cite{Premper2007}.
%According to the Rashba model, breaking inversion symmetry along the surface normal gives rise to in-plane spin polarization perpendicular to $k_{\parallel}$. In addition, in-plane asymmetries in the eslectrostatic potential can induce spin splitting accompanied by a finite out-of-plane ($z$) component of the spin polarization \cite{Premper2007}. 
The free-standing Bi$_3$Sb$_1$ layer exhibits sizable in-plane spin polarization (up to 21\%), which cannot be explained by the small vertical corrugation ($\sim$0.05~\AA). It is therefore attributed to Sb-induced asymmetries in the electronic potential, arising from the large difference in atomic number between Sb ($Z=50$) and Bi ($Z=83$). Notably, the resulting spin polarization remains perpendicular to $k{\parallel}$. Even more remarkable is the presence of a comparable spin polarization (up to 20\%) along the surface normal, as shown in Fig.~\ref{fig:SPINunsup}(d). The largest spin-polarization values occur in bands with predominant $p_{xy}$ character, consistent with their strong sensitivity to the in-plane potential.\cite{BiCu2alloy2020} These results for the free-standing Bi$_3$Sb$_1$ layer establish that isovalent substitution of heavier Bi atoms by lighter Sb ones is alone sufficient to induce significant asymmetries in the electronic potential, both along the surface normal and within the plane, giving rise to sizable spin splitting and spin polarization, without any significant accompanying structural or electronic reorganization.

\begin{figure}[ht]
\includegraphics[width=1.0 \columnwidth]{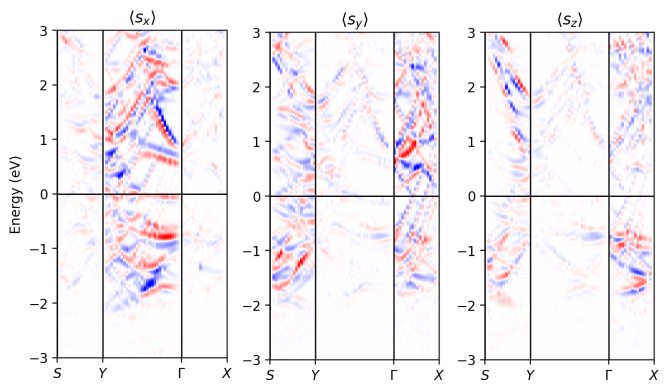}
\caption{(a)–(c) $k$-resolved spin projected density of states for the  $(3\times\sqrt{3})$–Bi$_3$Sb$_1$/Ag(111) system, along the three Cartesian directions, with the $\hat{z}$ direction corresponding to the surface normal. The magnitude and sign of the spin polarization are represented on a blue-white-red scale, where blue, white, and red correspond to -15\%, 0, and +15\%, respectively. }
  \label{fig:SPINsup}
\end{figure}

Finally, the interaction with the Ag(111) substrate significantly modifies the band structure of the free-standing Bi$_3$Sb$_1$ layer, with the $p_z$-derived states more strongly affected. Nevertheless, the main spin-splitting features identified in the unsupported layer are largely preserved in the supported system, as shown in Fig.~\ref{fig:SPINsup}. In particular, a sizable in-plane spin polarization along the $\hat{x}$ direction (up to 14\%) is observed along the $\bar{\Gamma}$–$\bar{Y}$ path, associated with Rashba-like split unoccupied bands, while the spin polarization along the surface normal also remains significant (up to 14\%) for the unoccupied bands along the $\bar{S}$–$\bar{Y}$ path. In contrast, when the overlayer–substrate interaction is included for the pure Bi$_4$ layer, it induces an in-plane spin polarization but no significant component along the surface normal (Fig.~S8 of the Supporting Material).

Although the predicted spin-splitting of the bands are expected to be present in the prepared surfaces, their direct observation lies beyond the resolution of a standard ARPES experiment, owing to the intrinsic broadening of the surface-state bands in Bi$_x$Sb$_{1-x}$/Ag(111). Their detection would require spin-resolved ARPES measurements.\cite{Gierz2011}

\section{Conclusions}

The electronic structure of single-layer Bi$_x$Sb$_{1-x}$ alloys grown on Ag(111) was investigated by angle-resolved photoemission spectroscopy (ARPES), complemented by density functional theory (DFT) calculations. Two samples with different stoichiometries were analyzed, with particular focus on the Bi-rich sample with composition close to Bi$_3$Sb$_1$.

ARPES measurements reveal four surface-state bands whose overall dispersion  depends only weakly on stoichiometry, as expected for an alloy formed by  isovalent elements. The good agreement between the experimental results and  the calculated band structures for both compositions supports the proposed  structural model, consisting of a rectangular $(3\times\sqrt{3})$  Bi$_x$Sb$_{1-x}$ overlayer with a four-atom unit cell.

An extended theoretical analysis provides further insight into the electronic structure of the system. First, the surface-state bands have predominantly $p_{xy}$ character, reflecting the weaker interaction of these orbitals with Ag bulk states compared with states of $p_z$ character, and indicating a substrate-induced orbital-filtering effect. Second, spin-resolved band-structure calculations reveal that Sb incorporation results in significant spin splitting and spin polarization, particularly along the surface normal. This is a consequence of the asymmetries in the electronic potential, both along the surface normal and within the plane, produced by the pressence of the Sb atoms.

By combining two isovalent elements, we isolate the effect of atomic mass on the spin texture: substituting heavier Bi atoms with lighter Sb ones is alone sufficient to induce significant asymmetries in the electronic potential, both along the surface normal and within the plane, giving rise to sizable spin splitting and spin polarization, without other significant structural or electronic band modifications. Although the intrinsic broadening of the surface-state bands in the present system prevents their experimental resolution, the spin-splitting of the bands are expected to be present in the Bi$_x$Sb$_{1-x}$/Ag(111) surface. Through a concrete model system, our work illustrates a general principle: that incorporating a lighter isovalent element can significantly increase spin polarization, and may thus provide a useful design guideline for understanding and engineering related systems.

\section{Supporting Information}

\section{Acknowledgments}

We acknowledge the financial support from CONICET (Grant PIP-2021-1404)

\bibliography{BiSb}

\end{document}